\documentclass[twocolumn,showpacs,preprintnumbers,amsmath,amssymb]{revtex4}
\usepackage{graphicx}
\usepackage{dcolumn}
\usepackage{bm}

 \newcommand{\pdf}{\textit{pdf}}
 \newcommand{\chf}{\textit{chf}}
 \newcommand{\lch}{\textit{lch}}

 \newcommand{\refeq}[1]{~(\ref{#1})}

\begin{document}

\title{L\'evy processes and Schr\"odinger equation}

\author{Nicola \surname{Cufaro Petroni}}
 \email{cufaro@ba.infn.it}
 \affiliation{Department of Mathematics and TIRES, Bari
 University;\\
 INFN Sezione di Bari, \\
via E Orabona 4, 70125 Bari, Italy}
\author{Modesto Pusterla}
 \email{pusterla@pd.infn.it}
 \affiliation{Department of Physics, Padova
 University;\\
 INFN Sezione di Padova, \\
via F.\ Marzolo 8, 35100 Padova, Italy}

\begin{abstract}
\noindent We analyze the extension of the well known relation
between Brownian motion and Schr\"odinger equation to the family of
the L\'evy processes. We consider a L\'evy--Schr\"odinger equation
where the usual kinetic energy operator -- the Laplacian -- is
generalized by means of a selfadjoint, pseudodifferential operator
whose symbol is the logarithmic characteristic of an infinitely
divisible law. The L\'evy--Khintchin formula shows then how to write
down this operator in an integro--differential form. When the
underlying L\'evy process is stable we recover as a particular case
the fractional Schr\" odinger equation. A few examples are finally
given and we find that there are physically relevant models (such as
a form of the relativistic Schr\"odinger equation) that are in the
domain of the non stable L\'evy--Schr\"odinger equations.
\end{abstract}


 \maketitle

\section{Introduction}\label{introduction}

That the Schr\"odinger equation can be linked to some underlying
stochastic process is well known since longtime. This idea has
received along the years a number of different formulations: from
the Feynman path integral~\cite{feynman}, through the Bohm--Vigier
model~\cite{bohmvigier}, to the Nelson stochastic
mechanics~\cite{nelson,guerra}. In all these models the underlying
stochastic process powering the random fluctuations is a Gaussian
Brownian motion, and the focus of the interest is the (non
relativistic) Schr\"odinger equation of quantum mechanics. This
particular choice is understandable because on the one hand the
Gaussian Brownian motion is the the most natural and widely explored
example of Markov process available, and on the other hand its
connection with the Schr\"odinger equation has always lent the hope
of a finer understanding of quantum mysteries.

In the framework of stochastic mechanics, however, this standpoint
can be considerably broadened since in fact this theory is a model
for systems more general than quantum mechanics: a general
\emph{dynamical theory of Brownian motion} that can be applied to
several physical problems~\cite{applications,cufaro,paul}. On the
other hand in recent years we have witnessed a considerable growth
of interest in non Gaussian stochastic processes, and in particular
in the L\'evy
processes~\cite{sato,protter,cont,applebaum,barndorff}. This is a
field that was initially explored in the 30's and 40's of last
century~\cite{levy,gnedenko,loeve}, but that achieved a full
blossoming of research only in the last decades, also as a
consequence of the tumultuous development of computing facilities.
This interest is witnessed by the large field of the possible
applications of these more general processes from statistical
mechanics~\cite{paul} to mathematical
finance~\cite{cont,bouchaud,mantegna}. In the physical field,
however, the research scope is presently rather confined to a
particular kind of L\'evy processes: the stable processes and the
corresponding fractional calculus~\cite{metzler,mainardi}, while in
the financial domain a vastly more general type of processes is at
present in use. For instance the possibility of widening the
perspective of the Schr\"odinger--Brownian pair has been considered
also recently~\cite{laskin}, but the Schr\"odinger equation has been
generalized only to a fractional Schr\"odinger equation. At our
knowledge, instead, the association of the more general L\'evy
infinitely divisible processes to the Schr\"odinger equation has
already been recognized as an important tool only once, in a
paper~\cite{garbaczewski} principally concerned with the analysis of
the relativistic Schr\"odinger equation in the framework of the so
called \emph{Schr\"odinger problem} solved by means of Bernstein
processes~\cite{bernstein}. Here, on the other hand, we intend to
develop the Nelson stochastic mechanics as a dynamical theory of the
infinitely divisible processes, and to widen the horizon of its
applications even to cases different from the quantum
systems~\cite{applications,cufaro}.

The appeal of the stable distributions is justified by the
properties of scaling and self--similarity displayed by the
corresponding processes, but it must also be remarked that these
distributions show a few features that partly impair their
usefulness as empirical models. First of all the non gaussian stable
laws always have infinite variance. This makes them rather suspect
as a realistic tool and prompts the introduction of
\textsl{truncated} stable distributions which, however, are no
longer stable. Then the range of the $x$ decay rates of the
probability density functions can not exceed $x^{-3}$, and this too
introduces a particular rigidity in these models. On the other hand
the more general L\'evy processes are generated by infinitely
divisible laws and do not necessarily show these disturbing
features, but they can be more difficult to analyze and to
simulate~\cite{heyde,cufaro07,berg}. Beside the fact that they do
not have natural scaling properties, the probability density
function of their increments could be explicitly known only at one
time instant. In fact, while their time evolution can always be
explicitly given in terms of characteristic functions, their
marginal densities may not be calculable. This is a feature,
however, that they share with most stable processes, since the
probability density functions of the non gaussian stable laws are
explicitly known only in precious few cases. On the other hand some
new applications in the physical domain for L\'evy, infinitely
divisible but not stable processes begin to emerge: in particular
the statistical characteristics of some recent model of the
collective motion in the charged particle accelerator beams seem to
point exactly in the direction of some kind of Student infinitely
divisible process~\cite{cufaro,vivoli}.

This paper is devoted to a discussion of a generalization of the
Schr\"odinger equation which takes into account the entire family of
the L\'evy processes: we will propose an equation where the
infinitesimal generator of the Brownian semigroup (the Laplacian) is
substituted by the more general generator of a L\'evy semigroup. As
it happens this will be a pseudodifferential operator (as, in
particular, in the fractional case), and the L\'evy--Khintchin
formula will give us the opportunity to write it down in the form of
an explicit integro--differential operator by putting in evidence
its continuous (Gaussian) and its jumping (non Gaussian) parts. The
advantages of this formulation are many: first of all the widening
of the increment laws from the stable to the infinitely divisible
case will offer the possibility of having realistic, finite
variances. Moreover both the possible presence of a Gaussian
component in the L\'evy--Khintchin formula, and the wide spectrum of
decay velocities of the increment probability densities will afford
the possibility of having models with differences from the usual
Brownian (and usual quantum mechanical, Schr\"odinger) case as small
as we want. In this sense we could speak of small corrections to the
quantum mechanical, Schr\"odinger equation. Last but not least,
there are examples of non stable L\'evy processes which are
connected to a particular form of the quantum, relativistic
Schr\"odinger equation: an important link that was missing in the
original Nelson model. It seems in fact that -- as was already
pointed out a few years ago~\cite{garbaczewski,deangelis} -- we can
only recover some kind of relativistic quantum mechanics if we widen
the field of the underlying stochastic processes at least to that of
the selfdecomposable L\'evy processes. To avoid formal complications
we will confine our discussion to the case of processes in just one
spatial dimension: generalizations will be straightforward.

\section{A heuristic discussion}

Let us start from the non relativistic, free Schr\"odinger equation
associated to its propagator or Green function $G(x,t|y,s)$ (see for
example~\cite{feynman})
\begin{eqnarray}
  i\hbar\partial_t\psi(x,t) &=& -\frac{\hbar^2}{2m}\,\partial_x^2\psi(x,t) \label{schrfree}\\
  G(x,t|y,s) &=& \frac{1}{\sqrt{2\pi
  i(t-s)\hbar/m}}\,e^{-\frac{(x-y)^2}{2i(t-s)\hbar/m}}\label{propagator} \\
  \psi(x,t)&=&\int_{-\infty}^{+\infty}G(x,t|y,s)\,\psi(y,s)\,dy\label{propagation}
\end{eqnarray}
and compare it with the Fokker--Planck equation of a Wiener process
(Brownian motion) with diffusion coefficient $D$, \pdf\ (probability
density function) $q(x,t)$ and transition \pdf\ $p(x,t|y,s)$ (see
for example~\cite{gardiner})
\begin{eqnarray}
  \partial_t q(x,t) &=& D\,\partial_x^2 q(x,t)\label{fpe} \\
  p(x,t|y,s) &=& \frac{1}{\sqrt{4\pi
  (t-s)D}}\,e^{-\frac{(x-y)^2}{4(t-s)D}}\label{transpdf} \\
  q(x,t)&=&\int_{-\infty}^{+\infty}p(x,t|y,s)\,q(y,s)\,dy\label{chk}
\end{eqnarray}
It is apparent that there is a simple, formal procedure transforming
the two structures one into the other:
\begin{equation}\label{substitution}
    D=\frac{\hbar}{2m}\,,\qquad\quad t\longleftrightarrow
    it\nonumber
\end{equation}
It is well known that this is just the result of an analytic
continuation in the complex plane. There are of course important
differences between $G$ and $p$. For example while $p$ and $q$ are
well behaved \pdf's, $G$ is not a wave function, as can be seen also
from a simple dimensional argument. This simple symmetry can then be
deceptive, but a better understanding of its true meaning can be
achieved either by means of the Feynman path integration with a free
Lagrangian of the usual quadratic form, or through the Madelung
decomposition~\cite{madelung} of\refeq{schrfree} and its subsequent
stochastic mechanical model~\cite{nelson,guerra}. Our aim here is to
generalize to distributions other than Gaussian this simple shortcut
from Wiener process to Schr\"odinger equation, and to analyze its
most immediate consequences. We postpone to a subsequent paper a
more detailed analysis in the framework of stochastic mechanics.

Let us see first of all what kind of role the gaussian distribution
plays in our Wiener--Schr\"odinger scheme. The \pdf\ and the \chf\
(characteristic function) of a Gaussian law $\mathcal{N}(0,a^2)$
\begin{equation*}
    q(x)=\frac{e^{-x^2/2a^2}}{\sqrt{2\pi a^2}}
    \,,\qquad\varphi(u)=e^{-a^2u^2/2}
\end{equation*}
satisfy the relations
\begin{eqnarray}\label{chf}
    \varphi(u)&=&\int_{-\infty}^{+\infty}q(x)\,e^{iux}\,dx\,,\label{chf}\\
    q(x)&=&\frac{1}{2\pi}\int_{-\infty}^{+\infty}\varphi(u)\,e^{-iux}\,du\,.\nonumber
\end{eqnarray}
Then formally the propagator\refeq{propagator} and the transition
\pdf\refeq{transpdf} respectively have as \chf's
\begin{equation*}
    e^{-iD(t-s)u^2}=\left[\varphi(u)\right]^{i(t-s)/\tau}\!\!,\quad
    e^{-D(t-s)u^2}=\left[\varphi(u)\right]^{(t-s)/\tau}
\end{equation*}
where now $\varphi(u)=e^{-D\tau u^2}=e^{-\tau\hbar u^2/2m}$ is the
\chf\ of a Gaussian law $\mathcal{N}(0,2D\tau)$, and $\tau$ is a
time constant introduced in order to have dimensionless exponents.
From\refeq{chf} we then have
\begin{eqnarray}
    G(x,t|y,s)\!\!&=&\!\!\frac{1}{2\pi}\int_{-\infty}^{+\infty}
    \left[\varphi(u)\right]^{i(t-s)/\tau}e^{-iu(x-y)}\,du\label{propagator1}\\
    p(x,t|y,s)\!\!&=&\!\!\frac{1}{2\pi}\int_{-\infty}^{+\infty}
    \left[\varphi(u)\right]^{(t-s)/\tau}e^{-iu(x-y)}\,du\label{transpdf1}
\end{eqnarray}
In both cases the starting point is the same \chf\ $\varphi$ of the
centered, normal law $\mathcal{N}(0\,,\,2D\tau)$. Then we consider
the \chf\ $[\varphi(u)]^{(t-s)/\tau}$ of the $t-s$ stationary
increments of the Wiener process, we pass to the imaginary time
variables ($t\leftrightarrow it$), and finally we get
from\refeq{propagator1} the Schr\"odinger
propagator\refeq{propagator}. The time scale $\tau$ incorporated in
the initial normal law disappears in the subsequent steps when we
generate the \chf\ of the increments. This feature is common to all
the stable laws and is the embodiment of the stable processes
selfsimilarity.

The equation\refeq{schrfree} and\refeq{fpe} can now be easily
deduced respectively from\refeq{propagator1} and\refeq{transpdf1}.
For instance from\refeq{propagation} and\refeq{propagator1} we have
\begin{equation*}
  \psi(x,t) = \int_{-\infty}^{+\infty}dy\,\frac{\psi(y,s)}{2\pi}
                 \int_{-\infty}^{+\infty}e^{-iDu^2(t-s)}e^{-iu(x-y)}du
\end{equation*}
and then we can write
\begin{eqnarray*}
  \lefteqn{i\partial_t\psi(x,t) =}\\
  && \int_{-\infty}^{+\infty}\!dy\,\frac{\psi(y,s)}{2\pi}
                 \int_{-\infty}^{+\infty}Du^2e^{-iDu^2(t-s)}e^{-iu(x-y)}du=\\
  && D\!\int_{-\infty}^{+\infty}\!\!dy\,\frac{\psi(y,s)}{2\pi}
                 \int_{-\infty}^{+\infty}\!\!(i\partial_x)^2e^{-iDu^2(t-s)}e^{-iu(x-y)}du=\\
  &&-D\,\partial_x^2\int_{-\infty}^{+\infty}\!dy\,\frac{\psi(y,s)}{2\pi}
                 \int_{-\infty}^{+\infty}\!\!e^{-iDu^2(t-s)}e^{-iu(x-y)}du=\\
  &&-D\,\partial_x^2\psi(x,t)
\end{eqnarray*}
which -- but for a factor $\hbar$ -- is the free, non relativistic
Schr\"odinger equation\refeq{schrfree}.

We are interested now in reproducing these well known steps starting
with the \chf\ of a non Gaussian law. Take now an infinitely
divisible -- in general non Gaussian -- law with \chf\ $\varphi(u)$,
and let $\eta(u)=\ln\varphi(u)$ be its \lch\ (logarithmic
characteristic). In the following we will restrict us to centered
laws, and we will justify this choice in the subsequent sections.
\emph{Infinite divisibility} essentially is the property of a \chf\
$\varphi$ which guarantees that also $\varphi^{t/\tau}$ is a
legitimate \chf\ for every real $t$. About the infinitely divisible
laws and their intimate relation with the L\'evy processes there is
a vast literature (see for example~\cite{sato,loeve}, and for a
short introduction~\cite{cufaro08}). The law of the increment of the
corresponding L\'evy process then is
$\left[\varphi(u)\right]^{(t-s)/\tau}$ and its transition \pdf\
is\refeq{transpdf1} with our -- possibly non Gaussian -- infinitely
divisible \chf. Then, following the procedure previously outlined
for the Wiener--Schr\"odinger equation, the wave function propagator
is\refeq{propagator1}  with our new $\varphi$, and hence
from\refeq{propagation} the time evolution is ruled by
\begin{equation*}
    \psi(x,t)=
    \int_{-\infty}^{+\infty}\!\!dy\,\frac{\psi(y,s)}{2\pi}\!
    \int_{-\infty}^{+\infty}\!\left[\varphi(u)\right]^{i(t-s)/\tau}e^{-iu(x-y)}du.
\end{equation*}
The differential equation can then be deduced as in the Gaussian
case and is
\begin{eqnarray}
    \lefteqn{i\partial_t\psi(x,t)=-\frac{1}{\tau}\,\eta(\partial_x)\psi(x,t)=}\label{newschr}\\
    &&\!\!\int_{-\infty}^{+\infty}\!\!\!dy\,\frac{\psi(y,s)}{2\pi}
                 \!\int_{-\infty}^{+\infty}\!\!\!-\frac{\eta(u)}{\tau}\,
                 \left[\varphi(u)\right]^{i(t-s)/\tau}\!e^{-iu(x-y)}du\nonumber
\end{eqnarray}
where now $\ln[\varphi(\partial_x)]=\eta(\partial_x)$ is a
\textit{pseudodifferential operator} with symbol $\eta(u)$ that is
defined through the use of the Fourier
transforms~\cite{cont,applebaum,taylor,jacob}. A pseudodifferential
operator $L$ on a suitable set of functions $h(x)$, is associated to
a function $\ell(u)$ called the \emph{symbol} of $L$, and operates
in the following way: if, coherently with the definition of the
\chf\ of a law (see~\cite{cont}), the Fourier transform of a
function $h$ is defined
\begin{equation}\label{ft}
    \widehat{h}(u)=\int_{-\infty}^{+\infty}h(x)\,e^{iux}dx
\end{equation}
then
\begin{equation}\label{pdop}
    (Lh)(x)=\frac{1}{2\pi}\int_{-\infty}^{+\infty}\ell(u)\,\widehat{h}(u)\,e^{-iux}du.
\end{equation}
When the symbol is a polynomial of degree $n$
\begin{equation*}
    \ell(u)=\sum_{k=1}^na_k(iu)^k
\end{equation*}
then $L$ is a simple differential operator of order $n$
\begin{equation*}
    L=\sum_{k=1}^na_k\partial_x^k
\end{equation*}
as can be easily seen from the properties of the Fourier transforms.
However, even if $\ell(u)$ is not a polynomial, equation\refeq{pdop}
defines an operator which is called pseudo\-differential. We will
now analyze the properties and the role of our pseudodifferential
operator $\eta(\partial_x)$ to see if\refeq{newschr} can reasonably
be considered as a generalized Schr\"odinger equation.

\section{Semigroups and generators}

Let $X(t)$ be a one dimensional L\'evy process, namely a process
with stationary and independent increments, and $X(0)=0$ almost
surely. The \chf\ of its increments on a time interval $\Delta t$
then is $[\varphi(u)]^{\Delta t/\tau}$ where $\varphi(u)$ is an
infinitely divisible law, and $\tau$ a time scale parameter (see for
example~\cite{sato} and~\cite{applebaum} for details about L\'evy
processes). It is well known that $\eta(u)=\ln\varphi(u)$ is the
\emph{lch} of an infinitely divisible law if and only if it
satisfies the L\'evy--Khintchin formula~\cite{applebaum}
\begin{equation}\label{lkformula}
    \eta(u)=i\gamma
    u-\frac{\beta^2}{2}\,u^2+\int_\mathbb{R}\left[e^{iux}-1-iuxI_{[-1,1]}(x)\right]\,\nu(dx)
\end{equation}
where $\gamma, \beta\in \mathbb{R}$, $I_A$ is the indicator 0--1
function of the set $A$, and $\nu(\,\cdot)$ is the L\'evy measure of
our infinitely divisible law, namely a measure on $\mathbb{R}$ such
that $\nu(\{0\})=0$ and
\begin{equation*}
    \int_\mathbb{R}(x^2\wedge1)\,\nu(dx)<+\infty\,.
\end{equation*}
The integrals involving the L\'evy measure $\nu$ should then in
general be calculated on $\mathbb{R}-\{0\}$ to take into account its
behavior near $y=0$. The triplet $(\gamma,\beta,\nu)$ completely
determines the L\'evy process and is also called its
\emph{characteristic triplet}. There are a few equivalent
formulations of this important result~\cite{sato}. In particular the
truncation function $I_{[-1,1]}(x)$ can be chosen in several
different ways; this choice, however, will affect only the value of
$\gamma$, while $\beta$ and $\nu$ would be left unchanged.

To every L\'evy process is associated a semigroup $(T_t)_{t\geq0}$
acting on the space $\mathcal{D}$ of the measurable, bounded
functions~\cite{applebaum}: if $f\in\mathcal{D}$, we have
$\left(T_tf\right)(x)=\mathbf{E}\left[f\left(X(t)+x\right)\right]$,
where $\mathbf{E}$ is the expectation. The infinitesimal generator
$A$ of the semigroup (see~\cite{applebaum} p.\ 131) is now defined
on the domain $\mathcal{D}_A$ of the functions $f\in\mathcal{D}$
such that the limit (in norm on $\mathcal{D}$)
\begin{equation*}
    Af=\lim_{t\rightarrow0^+}\frac{T_tf-f}{t}
\end{equation*}
exists. It can be proved (see~\cite{applebaum} p.\ 139) that the
generators of a L\'evy process are pseudodifferential operators that
can be extended to the Schwartz space $\mathcal{S}$ of the rapidly
decreasing functions. In particular we find that the symbol of $A$
is nothing else than $\eta(u)$, namely $A=\eta(\partial_x)$, and
that from the L\'evy--Khintchin formula\refeq{lkformula} we have
\begin{eqnarray*}
   \lefteqn{ [\eta(\partial_x)f](x)=\gamma(\partial_xf)(x)+\frac{\beta^2}{2}(\partial_x^2f)(x)}\\
    &&+\int_\mathbb{R}\left[f(x+y)-f(x)-y(\partial_xf)(x)I_{[-1,1]}(y)\right]\,\nu(dy)\,.
\end{eqnarray*}
In other words our pseudodifferential operator $\eta(\partial_x)$ of
equation\refeq{newschr} is the generator of the underlying L\'evy
process, and thanks to the L\'evy--Khintchin formula it also has an
explicit expression in terms of integro--differential operators.

The generator $A=\eta(\partial_x)$ can finally be extended to
$L^2(\mathbb{R})$ which is a Hilbert space, so that we can also
discuss its self-adjointness. In particular if $X(t)$ is a L\'evy
process, then its infinitesimal generator $A=\eta(\partial_y)$ will
be self-adjoint in $L^2(\mathbb{R})$ if and only if $X(t)$ is
centered and symmetric, namely if the symbol $\eta(u)$ is real with
\begin{equation}\label{lkformulas}
    \eta(u)=-\frac{\beta^2}{2}\,u^2+\int_\mathbb{R}(\cos
    ux-1)\,\nu(dx)
\end{equation}
where $\nu(\cdot\,)$ is a symmetric L\'evy measure
(see~\cite{applebaum} p.\ 154). A L\'evy measure is symmetric when
$\nu(B)=\nu(-B)$ for every Borel measurable set, where $-B=\{x;
-x\in B\}$. As a consequence the self-adjoint generators of the
centered and symmetric L\'evy processes enjoy the following
simplified, integro-differential form
\begin{eqnarray}
    (Af)(x)&=&[\eta(\partial_x)f](x)\label{generators}\\
    &=&\frac{\beta^2}{2}\left(\partial_x^2f\right)(x)+\int_\mathbb{R}[f(x+y)-f(x)]\,\nu(dy).\nonumber
\end{eqnarray}
It is also possible to show that $-\eta(\partial_x)$ is not only
self-adjoint, but also positive on $L^2(\mathbb{R})$ in the sense
that for every $f\in L^2(\mathbb{R})$ we have
$-(f,\eta(\partial_x)f)\geq0$, and this is equivalent to say that
the spectrum of $-\eta(\partial_x)$ lies entirely in $[0,+\infty)$.

We come back now to our problem of giving a generalized,
L\'evy--Schr\"odinger equation. Let $X(t)$ be a centered, symmetric
L\'evy process with \pdf\ and transition \pdf: we know that there is
a centered, infinitely divisible law with \chf\
$\varphi(u)=e^{\eta(u)}$ such that the \chf\ of its stationary
increments $\Delta X(t)=X(t+\Delta t)-X(t)$ is
\begin{equation*}
    [\varphi(u)]^{\Delta t/\tau}=e^{\eta(u)\Delta t/\tau}
\end{equation*}
for a suitable time scale parameter $\tau$. The transition \pdf\
then is\refeq{transpdf1}, so that by means of the substitution
$t\leftrightarrow it$ we get the propagator $G$ of
equation\refeq{propagator1}. A wave function $\psi$ then evolves
following\refeq{newschr}, and since our process is centered and
symmetric the generator $\eta(\partial_x)$ is self-adjoint and has
the integro--differential expression\refeq{generators}. That means
that our proposed equation takes the form
\begin{eqnarray}
    i\partial_t\psi(x,t)\!\!&=&\!\!-\frac{\eta(\partial_x)}{\tau}\psi(x,t)\label{newschr1}\\
                 \!\! &=&\!\! -\frac{\beta^2}{2\tau}\,\partial_x^2\psi(x)
                   -\frac{1}{\tau}\int_\mathbb{R}[\psi(x+y)-\psi(x)]\,\nu(dy).\nonumber
\end{eqnarray}
When $X(t)$ is a Gaussian Wiener process we know that
$\eta(u)=-\beta^2u^2/2$ and
$A=\eta(\partial_x)=\frac{\beta^2}{2}\partial_x^2$. Then the process
evolution equation is reduced to the Fokker--Planck
equation\refeq{fpe} with $D=\beta^2/2\tau$, and we have argued that
in this case the usual non relativistic, free Schr\"odinger equation
can de obtained by means of the substitution\refeq{substitution}. In
fact this amounts to take the -- self-adjoint and positive in
$L^2(\mathbb{R})$ -- (pseudo)differential operator
$-\hbar\eta(\partial_x)/\tau=-\hbar D\partial_x^2$ as the kinetic
energy operator. We propose here to extend this shortcut also to the
case of non Gaussian L\'evy processes. Apparently this is just a
formal analogy, but there are at least two ways to make it more
compelling: the Feynman path integral, and the Nelson stochastic
mechanics. Here we only take for granted this association between
generators and kinetic energy operators in order to establish the
form of the free L\'evy--Schr\"odinger equation and discuss its
first consequences; we instead postpone to a subsequent paper a more
rigorous derivation based on the application of Nelson stochastic
mechanics to L\'evy processes. We will remember however that the
method of Feynman path integrals has already been used in the
particular case of the fractional Schr\"odinger
equations~\cite{laskin} to obtain the same association, albeit in a
more restricted case: that of the stable laws, a particular class of
infinitely divisible laws. Note that we will adopt here only the
formal substitution $t\leftrightarrow it$, but we will not impose
$D=\hbar/2m$ because our generalizations are not necessarily
supposed to be some kind of quantum mechanics, but will rather
describe a \emph{dynamical theory of L\'evy processes} in the spirit
of Nelson stochastic mechanics. We can also introduce a constant
$\alpha$ with the dimensions of an action, so that our proposed free
L\'evy--Schr\"odinger equation\refeq{newschr1} becomes
\begin{eqnarray}\label{lseq}
    \lefteqn{i\alpha\partial_t\psi(x,t)=H_0\psi(x,t)=-\frac{\alpha}{\tau}\,\eta(\partial_x)\psi(x,t)=}\\
    &&-\alpha\frac{\beta^2}{2\tau}\,\partial_x^2\psi(x,t)
        -\frac{\alpha}{\tau}\int_\mathbb{R}[\psi(x+y,t)-\psi(x,t)]\,\nu(dy)\nonumber
\end{eqnarray}
where now the free hamiltonian operator $H_0$ has the dimensions of
an energy. This integro-differential hamiltonian $H_0$ is
self-adjoint and positive on $L^2(\mathbb{R})$ so that it is a good
kinetic energy operator. Everything that can be deduced about the
usual Schr\"odinger equation from the positivity and
self-adjointness of $H_0$ can also be of course extended to our
L\'evy--Schr\"odinger equation\refeq{lseq}. In particular the
conservation of the probability in the sense that, if $|\psi|^2$
plays the role of the position \pdf, then the norm $\|\psi\|^2$ will
be constant. We could finally add a potential $V(x)$ to\refeq{lseq}
and get a complete L\'evy--Schr\"odinger equation
\begin{eqnarray}\label{lseqpot}
    i\alpha\partial_t\psi(x,t)&=&H\psi(x,t)=\nonumber\\
    &&-\frac{\alpha}{\tau}\,\eta(\partial_x)\psi(x,t)+V(x)\psi(x,t)
\end{eqnarray}
where the hamiltonian is now $H=H_0+V$.

\section{Discussion and examples}

The free L\'evy--Schr\"odinger hamiltonian of equation\refeq{lseq}
contains two parts: the usual kinetic energy
$-\frac{\alpha\beta^2}{2\tau}\,\partial_x^2$ related to the Gaussian
part of the process; and the jump part which is given by means of an
integral with a symmetric L\'evy measure $\nu$. Of course, depending
on the nature of the underlying process, the equation\refeq{lseq}
can contain these components in different mixtures. If the
underlying process is purely Gaussian then the L\'evy measure $\nu$
vanishes and\refeq{lseq} is reduced to the usual Schr\"odinger
equation. On the other hand when the initial process is totally non
Gaussian, then $\beta=0$ and we get a pure jump
L\'evy--Schr\"odinger equation. In general both terms are present
and if for instance we introduce $\omega=1/\tau$ and choose
$\alpha=\hbar$ and $\beta^2=\alpha \tau/m$, then\refeq{lseq} takes
the form
\begin{eqnarray*}
    \lefteqn{i\hbar\partial_t\psi(x,t)=}\\
    &&-\frac{\hbar^2}{2m}\,\partial_x^2\psi(x,t)-\hbar\omega\int_\mathbb{R}[\psi(x+y,t)-\psi(x,t)]\,\nu(dy)\,.
\end{eqnarray*}
Here the jump term can be considered as a correction to the usual
Schr\"odinger equation and its weight, the energy $\hbar\omega$, is
at present a free parameter. In particular for $\beta=0$ we get pure
jump Schr\"odinger equations of the form
\begin{equation}\label{purejumpse}
    i\partial_t\psi(x,t)=
    -\omega\int_\mathbb{R}[\psi(x+y,t)-\psi(x,t)]\,\nu(dy)\,.
\end{equation}
Of course these remarks emphasize the fact that the explicit form of
the L\'evy--Schr\"odinger equation will depend on the choice of the
particular L\'evy measure $\nu$.

\subsection{Stationary free solutions}\label{stfreesol}
Let us consider first the stationary solutions of\refeq{lseq}:
taking
\begin{equation*}
    \psi(x,t)=e^{-iEt/\alpha}\phi(x)\,,\qquad\quad
    i\alpha\partial_t\psi(x,t)=E\psi(x,t)
\end{equation*}
we have that the spatial part $\phi(x)$ will be solution of
\begin{eqnarray}\label{statsol}
    H_0\phi(x)&=&-\frac{\alpha\beta^2}{2\tau}\,\phi''(x)
                   -\frac{\alpha}{\tau}\int_\mathbb{R}[\phi(x+y)-\phi(x)]\,\nu(dy)\nonumber\\
    &=&E\phi(x)
\end{eqnarray}
For the plane wave solutions
\begin{equation*}
    \phi(x)=e^{\pm iux}
\end{equation*}
and because of the symmetry of the L\'evy measure $\nu$,
equation\refeq{statsol} becomes
\begin{eqnarray*}
  E\phi(x) &=& \left[-\frac{\alpha\beta^2}{2\tau}\,u^2
                -\frac{\alpha}{\tau}\int_\mathbb{R}(e^{\pm iuy}-1)\,\nu(dy)\right]\phi(x) \\
   &=&\frac{\alpha}{\tau}\left[\frac{\beta^2u^2}{2}-\int_\mathbb{R}(\cos
   uy-1)\,\nu(dy)\right]\phi(x)\\
   &=&-\frac{\alpha}{\tau}\eta(u)\phi(x)
\end{eqnarray*}
and hence it is satisfied when between $E$ and $u$ the following
relation holds
\begin{equation*}
    E=-\frac{\alpha}{\tau}\eta(u)
\end{equation*}
Here $u$ is a wave number, while we are used to look for a relation
between energy $E$ and momentum $p$. If then we posit $p=\alpha u$
the energy--momentum relation for our free L\'evy--Schr\"odinger
equation is
\begin{equation}\label{energymomentum}
    E=-\frac{\alpha}{\tau}\,\eta\left(\frac{p}{\alpha}\right).
\end{equation}

\subsection{Some classes of infinitely divisible laws}
We will explore now a few examples of L\'evy--Schr\"odinger
equations associated to L\'evy processes. For a short summary of the
concepts used here see for example~\cite{cufaro08}. We will consider
the \chf's, L\'evy measures and infinitesimal generators of
centered, symmetric, infinitely divisible laws so
that\refeq{lkformulas} and\refeq{generators} hold. The form of the
L\'evy measure $\nu$ is then instrumental to explicitly show how the
pseudo-differential generator $\eta(\partial_x)$ works. It would
then be useful to list several classes of infinitely divisible laws
in a growing order of generality:
\begin{enumerate}
    \item \emph{Stable laws}: here we have~\cite{sato,cont}
    \begin{equation}\label{stable}
        \eta(u)=-\frac{(a|u|)^\lambda}{\lambda}\,;\qquad\quad0<\lambda\leq2\,,
    \end{equation}
    with the important particular cases
    \begin{equation*}
        \eta(u)=\left\{
           \begin{array}{ll}
             -a^2u^2/2 & \hbox{Gauss law ($\lambda=2$);} \\
             -a|u| & \hbox{Cauchy law ($\lambda=1$).}
           \end{array}
         \right.
    \end{equation*}
    Stable laws are selfdecomposable and hence their L\'evy measures
    are absolutely continuous~\cite{sato,cont} so that $\nu(dx)=W(x)\,dx$ with
    \begin{eqnarray}\label{stablelm}
        W(x)&=&\frac{B}{|x|^{\lambda+1}}\,,\\
        B&=&\left\{
          \begin{array}{ll}
            0 & \hbox{$\lambda=2$;} \\
            a/\pi & \hbox{$\lambda=1$;} \\
            \frac{-a^\lambda}{2\lambda\,\cos\frac{\lambda\pi}{2}\,\Gamma(-\lambda)} & \hbox{$\lambda\neq1,2$.}
          \end{array}
        \right.\nonumber
    \end{eqnarray}
    The infinitesimal generator, which in the Gauss case ($\lambda=2$) simply is
    \begin{equation*}
        \frac{a^2}{2}(\partial_x^2f)(x)\,,
    \end{equation*}
    for $0<\lambda<2$ becomes the pseudo-differential operator
    \begin{eqnarray*}
        \qquad [\eta(\partial_x)f](x)&=&-\frac{1}{\sqrt{2\pi}}\int_{-\infty}^{+\infty}\frac{(a|u|)^\lambda}{\lambda}
                     \,e^{iux}\widehat{f}\,(u)\,du\\
        &=&B\int_{-\infty}^{+\infty}\frac{f(x+y)-f(x)}{|y|^{\lambda+1}}\,dy
    \end{eqnarray*}
    which can also be expressed symbolically in terms of the \emph{fractional
    derivatives}~\cite{mainardi}
    \begin{eqnarray*}
        \qquad(\partial^\lambda_xf)(x)&=&-\frac{1}{\sqrt{2\pi}}\int_{-\infty}^{+\infty}|u|^\lambda
        e^{iux}\widehat{f}\,(u)\,du\\
        &=&-\frac{1}{\sqrt{2\pi}}\int_{-\infty}^{+\infty}(-\partial^2_x)^{\lambda/2}
        e^{iux}\widehat{f}\,(u)\,du\\
        &=&\left[-(-\partial^2_x)^{\lambda/2}f\right](x)\,.
    \end{eqnarray*}
    The Cauchy case is discussed at length
    in~\cite{garbaczewski}.

    \item \emph{Selfdecomposable (non stable) laws}: they are an important sub--family of infinitely divisible
    laws. Two examples are (for details and other examples
    see~\cite{cufaro07,cufaro08}) the Variance Gamma family
    (VG, $\lambda>0$) and a law connected to the relativistic quantum
    mechanics:
    \begin{equation*}
        \qquad\eta(u)=\left\{
           \begin{array}{ll}
             -\lambda\,\ln(1+a^2u^2), & \hbox{VG;} \\
             1-\sqrt{1+a^2u^2}, & \hbox{relativistic q.m.}
           \end{array}
         \right.
    \end{equation*}
    which have no Gaussian part ($\beta=0$ in the
    L\'evy--Khintchin formula) and produce pure jump processes.
    Their L\'evy measures have densities~\cite{cufaro07,cufaro08}
    \begin{equation*}
        \qquad W(x)=\left\{
          \begin{array}{ll}
            \lambda|x|^{-1}e^{-|x|/a}, & \hbox{VG;} \\
            (\pi|x|)^{-1}K_1(|x|/a), & \hbox{relativistic q.m.}
          \end{array}
        \right.
    \end{equation*}
    where $K_\lambda(z)$ is a modified Bessel function.
    Remark that, while for the VG \pdf\ can be explicitly given
    \begin{equation*}
       \qquad q(x)=\frac{2}{a\,2^\lambda\Gamma(\lambda)\sqrt{2\pi}}\left(\frac{|x|}{a}\right)^{\lambda-\frac{1}{2}}
               K_{\lambda-\frac{1}{2}}\left(\frac{|x|}{a}\right)
    \end{equation*}
    we have no elementary expressions for the
    Relativistic q.m.\ \pdf\
    \begin{equation*}
        q(x)=\frac{1}{2\pi}\int_{-\infty}^{+\infty}e^{1-\sqrt{1+a^2u^2}}e^{iux}du
    \end{equation*}
    which can then be calculated only numerically.

    \item \emph{Infinitely divisible (non selfdecomposable) laws}:
    The classical example of an infinitely divisible, non
    selfdecomposable law is the Poisson law of intensity $\lambda$, but the
    corresponding L\'evy process would not be symmetric. If however
    we take the \chf\ $\chi(u)$ of a centered, symmetric law, then the
    corresponding \textit{compound} Poisson process will be centered and symmetric with \lch
    \begin{equation*}
        \eta(u)=\lambda\left[\chi(u)-1\right]\,.
    \end{equation*}
    In the analysis of the corresponding L\'evy measure we must
    remember that now we can no longer take for granted that $\nu$
    is absolutely continuous. For example if the jump size can take only two
    values $\pm a$ ($a>0$) with equal probabilities $1/2$, then
    the \chf\ $\chi(u)=\cos au$ has no \pdf, $\eta(u)=\lambda(\cos
    au-1)$, and $\nu(dx)=\lambda F(dx)$ where the cumulative distribution
    \begin{equation}\label{symmetriccp}
        F(x)=\frac{\Theta(x-a)+\Theta(x+a)}{2}
    \end{equation}
    is a symmetric, two--steps function,
    and $\Theta(x)$ is the 0--1 Heaviside function. If on the other hand
    $\chi$ is a law with a \pdf\ $g(x)$, it is possible to show that also the L\'evy measure
    $\nu$ is absolutely continuous with a density
    \begin{equation*}
        W(x)=\lambda g(x)\,.
    \end{equation*}
    This completely specify the associated L\'evy process on the
    basis of the Poisson intensity $\lambda$, and of the law
    of the jump sizes.
\end{enumerate}

\subsection{Examples of L\'evy--Schr\"odinger equations}
We can now analyze a few examples of L\'evy--Schr\"odinger equations
based on the laws listed above. The first two cases have an apparent
physical meaning, while at present the other three (Variance--gamma,
stable and compound Poisson laws) are short of an immediate
interpretation.
\begin{enumerate}
    \item \emph{Non relativistic, quantum, free particle}: this is
    the well known case of the Gaussian Wiener process with
    $\eta(u)=-\beta^2u^2/2$ giving rise to the usual Schr\"odinger equation\refeq{schrfree}
    for a suitable identification of the parameters. In this case the energy--momentum
    relation\refeq{energymomentum} is
    \begin{equation*}
        E=-\frac{\alpha}{\tau}\left(-\frac{\beta^2}{2}\,\frac{p^2}{\alpha^2}\right)=
        \frac{\beta^2}{2\alpha \tau}\,p^2
    \end{equation*}
    and with $\alpha=\hbar$ and $\beta^2=\alpha
    \tau/m$ we get as usual
    \begin{equation*}
        E=\frac{p^2}{2m}\,,\qquad\quad p=\hbar u\,.
    \end{equation*}

    \item \emph{Relativistic, quantum, free particle}: it is
    interesting to remark at this point that there is a L\'evy
    process which is connected to the relativistic Schr\"odinger
    equation, in the same way as the Wiener process is connected to
    the non relativistic Schr\"odinger equation. Take the non
    stable, selfdecomposable law $\eta(u)=1-\sqrt{1+a^2u^2}$, and
    use the following identifications
    \begin{equation*}
        \frac{\alpha}{\tau}=mc^2\,,\qquad\quad
        a=\frac{\hbar}{mc}\,,\qquad\quad p=\hbar u
    \end{equation*}
    to find from\refeq{energymomentum}
    \begin{equation}\label{relativisticenergy}
        E=-mc^2\eta(u)=\sqrt{m^2c^4+p^2c^2}-mc^2
    \end{equation}
    which is the relativistic total energy less the rest energy
    $mc^2$: namely the kinetic energy. The corresponding L\'evy--Schr\"odinger equation is now
    \begin{equation*}
        \qquad i\hbar\partial_t\psi(x,t)=\left[\sqrt{m^2c^4-c^2\hbar^2\partial_x^2}-mc^2\right]\psi(x,t)
    \end{equation*}
    and is discussed at length in~\cite{garbaczewski}.
    Since the constant $-mc^2$ can be reabsorbed by means of a phase
    factor $e^{imc^2t/\hbar}$, the wave equation finally is
    \begin{equation}\label{relativisticse}
        i\hbar\partial_t\psi(x,t)=\sqrt{m^2c^4-c^2\hbar^2\partial_x^2}\,\psi(x,t)
    \end{equation}
    which is the simplest form of a relativistic, free Schr\"odinger
    equation~\cite{bjorken}.
    It is interesting to note that the L\'evy process behind the
    relativistic equation\refeq{relativisticse} is a pure jump
    process with an absolutely continuous L\'evy measure with \pdf
    \begin{equation*}
       \quad W(x)=\frac{1}{\pi|x|}\,K_1\left(\frac{|x|}{a}\right)=\frac{1}{\pi|x|}\,K_1\left(\frac{mc|x|}{\hbar}\right)
    \end{equation*}
    so that equation\refeq{relativisticse} can also be written as
    \begin{eqnarray}
        \qquad\lefteqn{i\hbar\partial_t\psi(x,t)=}\\
        &&-mc^2\int_\mathbb{R}\frac{\psi(x+y,t)-\psi(x,t)}{\pi|y|}
                       \,K_1\left(\frac{mc|y|}{\hbar}\right)\,dy\nonumber
    \end{eqnarray}
    From the form of the relativistic
    energy\refeq{relativisticenergy} also the usual
    relativistic corrections to the classical energy--momentum
    relation for small values of $p/c$ follow:
    \begin{eqnarray*}
        E&=&mc^2\left(\sqrt{1+\frac{p^2}{m^2c^2}}-1\right)\\
        &=&\frac{p^2}{2m}-\frac{p^4}{8m^3c^2}+o(p^5).
    \end{eqnarray*}
    Finally if we consider $E=H(p)$ as the Hamiltonian function of a
    relativistic free particle from the Hamilton equations we get
    \begin{equation*}
      \dot{q} = \partial_pH=\frac{p}{m}\,\frac{1}{\sqrt{1+p^2/m^2c^2}}
    \end{equation*}
    and here too we can see the relativistic correction to the
    classical kinematic relation $p=m\dot{q}$.

    \item \emph{Variance--Gamma laws}: for $\lambda=\frac{1}{2}$ we have
    \begin{eqnarray*}
        \quad\qquad\eta(u) = -\frac{1}{2}\ln(1+a^2u^2)\,,& & \!\!\!\!q(x) =
        \frac{1}{a\pi}\,K_0\left(\frac{|x|}{a}\right),\\
        W(x) &=& \frac{e^{-|x|/a}}{2|x|}\,.
    \end{eqnarray*}
    The Variance--Gamma processes are pure jump processes with no Gaussian part in the L\'evy--Khintchin
    formula\refeq{lkformula} ($\beta=0$), so that the
    L\'evy--Schr\"odinger equation becomes
    \begin{eqnarray}
        \lefteqn{i\alpha\partial_t\psi(x,t)=}\nonumber\\
        &&-\frac{\lambda\alpha}{\tau}
             \int_\mathbb{R}\frac{\psi(x+y,t)-\psi(x,t)}{|y|}\,e^{-|y|/a}dy
    \end{eqnarray}
   By choosing
   \begin{equation}\label{identifications}
        \alpha=\frac{ma^2}{\tau}\,,\qquad\quad p=\frac{ma^2}{\tau}\,u
   \end{equation}
   we have the following energy--momentum relation (for
   $p\tau/ma\to0$)
   \begin{eqnarray*}
        E &=&
        \frac{ma^2}{2\tau^2}\,\ln\left(1+\frac{\tau^2p^2}{m^2a^2}\right)\\
              &=&\frac{p^2}{2m}-\frac{\tau^2}{2ma^2}\,\frac{p^4}{m^2}+o(p^5)
   \end{eqnarray*}
   while with the identification $E=H(p)$ we can also recover the
   kinematic relations between $p$ and $\dot{q}$:
   \begin{equation*}
        \dot{q} = \frac{p/m}{1+\frac{\tau^2p^2}{a^2m^2}}
        =\frac{p}{m}-\frac{\tau^2p^3}{a^2m^3}+o(p^{4})
   \end{equation*}
   It is apparent then that again these equations are corrections to the
   classical relations.

    \item \emph{Stable laws}: for the non Gaussian stable laws we
    have the \lch\refeq{stable} and the L\'evy measure \pdf\refeq{stablelm}
    where $0<\lambda<2$. With a couple of
    exception (Cauchy and L\'evy laws), however, there are no
    elementary formulas for their \pdf's $q(x)$. Their
    L\'evy--Schr\"odinger equation is
    \begin{eqnarray}
        \qquad\lefteqn{i\alpha\partial_t\psi(x,t)=}\\
        &&\frac{\alpha}{\tau}\,
        \frac{a^\lambda}{2\lambda\cos\frac{\lambda\pi}{2}\Gamma(-\lambda)}\,
        \int_\mathbb{R}\frac{\psi(x+y,t)-\psi(x,t)}{|y|^{\lambda+1}}\,dy\nonumber
    \end{eqnarray}
    and with the identifications\refeq{identifications} the
    energy--momentum relations become
    \begin{equation*}\label{stableenergy}
       \qquad E=\frac{\alpha}{\tau}\,\frac{(a|u|)^\lambda}{\lambda}
        =\frac{2^{\lambda/2}}{\lambda}\,\left(\frac{ma^2}{\tau^2}\right)^{1-\lambda/2}\left(\frac{p^2}{2m}\right)^{\lambda/2}
    \end{equation*}
    This however looks not as a correction of the classical formula
    as in the other cases considered, but rather as a completely
    different formula. The same can be said of the kinematic
    relations between $\dot{q}$ and $p$ which are now
    \begin{equation*}\label{stablekinematics}
        \dot{q}=\frac{p}{m}\,\left(\frac{p^2\tau^2}{m^2a^2}\right)^{\lambda/2-1}
    \end{equation*}
    These relations are still another reason to consider not advisable to
    restrict an inquiry on L\'evy processes and Schr\"odinger
    equation only to the family of stable processes.

    \item\emph{Compound Poisson process}: for the symmetric,
    compound Poisson process with the L\'evy
    measure given by\refeq{symmetriccp} the pure jump L\'evy--Schr\"odinger
    equation\refeq{purejumpse} greatly simplifies as
    \begin{eqnarray*}
        \quad\lefteqn{i\partial_t\psi(x,t)=}\\
        &&-\frac{\lambda\omega}{2}\,\left[\psi(x+a,t)-2\psi(x,t)+\psi(x-a,t)\right]\,.
    \end{eqnarray*}
    We can now show that the usual Schr\"odinger equation\refeq{schrfree} can always be recovered
    as a limit case of this Poisson--Schr\"odinger equation. If $\psi$ is twice differentiable in $x$, we know that for
    $a\to0^+$
    \begin{eqnarray*}
        \lefteqn{\psi(x\pm a,t)=}\\
        &&\psi(x,t)\pm
        a\partial_x)\psi(x,t)+\frac{a^2}{2}\partial_x^2\psi(x,t)+o(a^2)
    \end{eqnarray*}
    and hence
    \begin{equation*}
        i\partial_t\psi(x,t)=-\frac{\lambda\omega a^2}{2}\,\partial_x^2\psi(x,t)+\lambda \,o(a^2)\,.
    \end{equation*}
    Now, if, as $a\to0^+$, also $\lambda\to+\infty$ in such a way
    that $\lambda a^2\to b^2$, then we have
    \begin{equation*}
        i\partial_t\psi(x,t)=-\frac{\omega b^2}{2}\,\partial_x^2\psi(x,t)
    \end{equation*}
    where $\omega b^2/2$ has the dimensions of a diffusion
    coefficient: namely, in the limit, we get a
    Wiener--Schr\"odinger equation of the type\refeq{schrfree}. This
    procedure can also be used to introduce small corrections in the
    coefficient $\hbar^2/2m$ of a Schr\"odinger equation.

\end{enumerate}

\subsection{Perfectly rigid walls}
An example of solution of the complete L\'evy--Schr\"odinger
equation\refeq{lseqpot} can easily be obtained in the case of a
system confined between two perfectly rigid walls symmetrically
located at $x=\pm L/2$. The discussion is similar to that of
Section~\ref{stfreesol}, but for the boundary conditions which now
require that the solutions vanish at $x=\pm L/2$. As a consequence
the solutions are the usual trigonometric functions with discrete
eigenvalues
\begin{equation*}
    E_n=-\frac{\alpha}{\tau}\,\eta(u_n)\,,\qquad\quad
    u_n=\frac{n\pi}{L}\,,\qquad n=1,2,\ldots
\end{equation*}
The form of the eigenvalue sequence will depend on the \lch\ $\eta$.
If the underlying process is a Wiener process we have
$\eta(u)=-\beta^2u^2/2$, and hence
\begin{equation*}
    E_n=\frac{\alpha\beta^2\pi^2}{2\tau L^2}\,n^2
\end{equation*}
which, with the identifications $\alpha=\hbar$ and
$\beta^2=\alpha\tau/m$, coincides with the usual quantum mechanical
result. On the other hand for an underlying Variance--Gamma noise we
find
\begin{equation*}
    E_n=\frac{\lambda\alpha}{\tau}\ln\left(1+\frac{a^2\pi^2}{L^2}\,n^2\right).
\end{equation*}
Finally for a symmetric, compound Poisson noise with intensity
$\lambda$ and equiprobable jump sizes $\pm a$ we get
\begin{equation*}
    E_n=\frac{\lambda\alpha}{\tau}\left(1-\cos \frac{a\pi n}{L}\right)
\end{equation*}
Apparently this no longer is a monotone increasing sequence: $E_n$
goes up and down between $0$ and $2\alpha\lambda/\tau$. If $a/L$ is
rational the sequence is periodic; on the other hand when $a/L$ is
irrational there are no coincident eigenvalues in the sequence, so
that the $E_n$ will fill the bandwidth between $0$ and
$2\alpha\lambda/\tau$.

\section{Conclusions}

We have discussed the possibility of generalizing the relation
between Brownian motion and Schr\"odinger equation by formally
associating the kinetic energy of a more general system to the
generator of a symmetric L\'evy process, namely to a
pseudodifferential operator whose symbol is the \lch\ of an
infinitely divisible law. This amounts to suppose, then, that this
new L\'evy--Schr\"odinger equation is based on an underlying L\'evy
process that can have both Gaussian and jumping components.

In recent years other extensions of the Schr\"odinger equation have
been put forward in the same spirit of our L\'evy--Schr\"odinger
equation. In particular we refer to several papers about fractional
Schr\"odinger equations~\cite{laskin} that explored the use of
fractional calculus in a kind of generalized quantum mechanics. As
it is clear from the previous sections, however, this is the
particular case when our underlying process is stable. The extension
to the infinitely divisible, non stable processes, on the other
hand, is meaningful because there are significant cases that are now
in the domain of our L\'evy--Schr\"odinger picture. In particular,
as shown also in~\cite{garbaczewski}, the simplest form of a
relativistic, free Schr\"odinger equation can be deduced from a
particular type of selfdecomposable, non stable process. Moreover in
many instances the new energy--momentum relations can be seen as
corrections to the classical relations for small values of certain
parameters. It must also be remembered that our model is not tied to
the use of processes with infinite variance: the variances can be
chosen to be finite even in a purely non Gaussian model -- as in the
case of the relativistic, free Schr\"odinger equation -- and can
then be used as a measure of the dispersion.

It is important now to explicitly give in full detail a derivation
of the L\'evy--Schr\"odinger equation from either the Feynman
integrals or a generalized stochastic mechanics. This seems to be
possible because the techniques of the stochastic calculus applied
to L\'evy processes are today in full
development~\cite{sato,protter,cont,applebaum,barndorff}, and at our
knowledge there is no apparent, fundamental impediment along this
road. At present our approach lacks this rigorous discussion of how
the L\'evy--Schr\"odinger equation comes out from the evolution
equations of the L\'evy processes. We have confined ourselves to
give only a few heuristic arguments based on both the identification
of the process generators as the kinetic energy operators, and the
analytic continuation of the time variable $t$ to its imaginary
counterpart $it$. We hinted, however, to two possible ways of giving
a more rigorous derivation: we can first of all follow the Feynman
integral road. In this case we should bear in mind that the
relations among kinetic energy and momentum are no longer the usual
relations: this is important to correctly write the Lagrangian in
the Feynman integral. Alternatively we can try to generalize Nelson
stochastic mechanics by adding a suitable dynamics to our L\'evy
processes, and this will be the subject of a future paper.

\begin{acknowledgments}
The authors want to thank S De Martino, S De Siena and F Illuminati
for useful discussions and suggestions, and for the long
collaboration which was -- and is -- instrumental in this enquiry.
\end{acknowledgments}

\vfill\eject

\end{document}